\documentclass[10pt,twocolumn]{article}

\usepackage[T1]{fontenc}
\usepackage{lmodern} 

\usepackage{amsmath,amssymb,amsthm}

\usepackage{graphicx}
\usepackage{float}
\usepackage{booktabs}
\usepackage{multirow}
\usepackage{tabularx}
\usepackage{array}
\newcolumntype{Y}{>{\raggedright\arraybackslash}X}
\newcolumntype{Z}{>{\centering\arraybackslash}X}
\newcolumntype{W}{>{\raggedleft\arraybackslash}X}

\usepackage{subcaption}

\usepackage{enumitem}
\setlist{topsep=3pt, itemsep=3pt, parsep=0pt, partopsep=0pt}

\setlist[enumerate,1]{label=\alph*), leftmargin=*}

\usepackage[numbers,sort&compress]{natbib}

\usepackage{lineno}

\usepackage[colorlinks=true,
linkcolor=blue,
citecolor=blue,
urlcolor=blue]{hyperref}

\usepackage{geometry}
\newtheorem{theorem}{Theorem}

\usepackage{titlesec}
\titleformat{\section}
{\normalsize\bfseries}
{\thesection.}
{0.5em}
{}
\titlespacing*{\section}{0pt}{2.5ex plus 0.2ex minus 0.2ex}{1.0ex}

\titleformat{\subsection}
{\normalsize\itshape}
{\thesubsection}
{0.5em}
{}
\titlespacing*{\subsection}{0pt}{1.5ex plus 0.2ex minus 0.2ex}{0.8ex}

\titleformat{\subsubsection}
{\normalsize\itshape}
{\thesubsubsection}
{0.6em}
{}
\titlespacing*{\subsubsection}{0pt}{1.2ex plus 0.2ex minus 0.2ex}{0.6ex}

\titleformat{\paragraph}
{\normalsize\bfseries}
{}
{0em}
{}

\titlespacing*{\paragraph}
{0pt}        
{1.2ex}      
{0.2em}      

\usepackage[font=small,labelfont=bf]{caption}
\DeclareMathOperator{\I}{I}

\newcommand{\USL}{\mathrm{USL}}
\newcommand{\LSL}{\mathrm{LSL}}

\usepackage[dvipsnames]{xcolor}
\usepackage[normalem]{ulem}
\usepackage{soul}
\sethlcolor{yellow!35}

\soulregister\cite7
\soulregister\citet7
\soulregister\citep7
\soulregister\ref7
\soulregister\pageref7

\usepackage{framed}
\definecolor{shadecolor}{rgb}{1,1,0.65}
\definecolor{mygrey}{gray}{0.965}

\usepackage{fancyhdr}
\title{Risk-Calibrated Process Capability Approval with Finite Samples}

\author{
	Fei Jiang\thanks{
		Independent Researcher, Seattle, WA, USA. 
		Correspondence to: jiangfeicq@gmail.com.
		Fei Jiang and Lei Yang contributed equally to this work.
	}
	\and
	Lei Yang\footnotemark[1]
}

\date{}
\begin{document}
\maketitle
\thispagestyle{fancy}
\pagestyle{plain} 
\begin{abstract}
	Process capability indices such as $C_{pk}$ are widely used in manufacturing to support supplier qualification, pilot-build release, and production approval. In practice, approval decisions are often based on deterministic threshold rules of the form $\widehat{C}_{pk} \ge C_0$. Because $\widehat{C}_{pk}$ is estimated from finite samples, however, such decisions are inherently stochastic, especially when the true capability lies near the approval threshold. This paper develops a risk-calibrated decision framework for process capability approval that explicitly accounts for estimation uncertainty and asymmetric operational loss. Capability approval is formulated as a binary statistical decision problem, leading to a rule of the form $\widehat{C}_{pk} \ge C_0 + k\,SE(\widehat{C}_{pk})$, where the calibration constant $k$ is determined either by a tolerable failure probability or by a false-accept/false-reject cost ratio. The resulting formulation unifies several commonly used procedures, including deterministic thresholding, lower confidence bound rules, and probability-based approval rules, and naturally extends them to cost-sensitive decision rules derived from asymmetric operational loss. Simulation experiments and an industrial case study show that risk calibration primarily affects near-threshold decisions, improves approval stability, and can substantially reduce expected operational loss when false acceptance is more costly than false rejection.
\end{abstract}


\setcounter{tocdepth}{2}  
\section{Introduction}

Process capability indices (PCIs) such as $C_p$, $C_{pk}$, $P_p$, and $P_{pk}$ are widely used in manufacturing quality engineering to assess whether a production process is capable of meeting engineering specifications. Capability analysis plays an important role in supplier qualification, pilot-build validation, and production release decisions \cite{juran1979quality,montgomery2020introduction,ISO22514-1-2014,ISO22514-4-2016}. Since the early introduction of process capability indices, extensive research has been devoted to their definition, interpretation, and extension under different process conditions \cite{kane1986process,chan1988new,boyles1991taguchi,vannman1995unified,kotz2002process,anis2008basic,wu2009overview}. Numerous variants of capability indices have been proposed to address practical issues such as off-target processes, asymmetric tolerances, and non-normal distributions \cite{chen1998incapability,chen2001capability,abbasi2016class,chen1997application,chen2001new,kovarik2014process}.

In industrial practice, capability approval decisions are often implemented using a simple deterministic threshold rule of the form $\widehat{C}_{pk} \ge C_0$, where $\widehat{C}_{pk}$ denotes the capability estimate computed from a validation sample and $C_0$ is a prescribed requirement, typically taking values such as $1.00$, $1.33$, or $1.67$. Processes with $\widehat{C}_{pk} \ge C_0$ are approved, whereas those below the threshold are rejected. Despite its simplicity, this deterministic threshold rule remains widely used in manufacturing quality practice.

Despite its widespread use, this rule conceals an important statistical feature: the estimated capability $\widehat{C}_{pk}$ is a random quantity rather than a fixed process characteristic. In many manufacturing applications, capability evaluation is based on moderate validation samples, often with sizes between $20$ and $50$, so sampling variability can cause the estimated capability to cross the approval threshold even when the true capability is close to the required threshold \cite{mahmoud2010estimating,alvarez2015methodological,jiang2026practical}.

The statistical properties of capability estimators and their associated inferential procedures have been studied extensively in the literature. Previous work has investigated the sampling distributions of capability indices, interval estimation, confidence bounds, and hypothesis testing procedures for capability requirements \cite{zhang1990interval,kushler1992confidence,pearn1992distributional,collins1995bootstrap,mathew2007generalized,pearn2004testing}. Lower confidence bound (LCB) approaches have been proposed for capability-based approval, including settings involving asymmetric tolerances and measurement uncertainty \cite{chang2008assessing,grau2011lower}. Bayesian approaches to capability testing have also been explored \cite{kargar2014bayesian}. While these contributions provide valuable inferential tools for capability analysis, they generally focus on estimation and testing rather than on the decision-theoretic structure of capability approval itself.

Recent theoretical results in \cite{jiang2026finite} highlight an important consequence of finite-sample uncertainty in threshold-based capability decisions. Specifically, they showed that when the true capability equals the approval threshold, the probability of approval converges asymptotically to $0.5$. More generally, when the true capability lies within an $O(n^{-1/2})$ neighborhood of the threshold, the approval probability remains nondegenerate and is governed by a scaled signal-to-noise ratio involving the distance from the threshold. These results imply that the conventional rule $\widehat{C}_{pk} \ge C_0$ should not be interpreted as a deterministic classification rule, but rather as an implicit statistical decision rule with nontrivial misclassification risk near the approval boundary.

This observation raises a natural question: how should capability approval decisions be designed when estimation uncertainty and the operational consequences of decision errors are taken into account explicitly?

In practical manufacturing environments, the two types of decision error often have very different consequences. Approving a process that is truly incapable may lead to product failures, warranty costs, recalls, or contractual violations. Conversely, rejecting a process that is actually capable may delay product launch, trigger unnecessary engineering investigations, or reduce production throughput. Such asymmetric consequences are consistent with the broader literature on statistical decision theory and conformity assessment, where measurement uncertainty must be explicitly incorporated into acceptance decisions \cite{berger2013statistical,pendrill2014using,ISO14253-1-2013}.

Motivated by this perspective, this paper develops a risk-calibrated framework for process capability approval under finite samples. Rather than treating capability approval as a point-estimation problem followed by ad hoc thresholding, we formulate it as a binary statistical decision problem under asymmetric loss. Under a normal approximation to the sampling distribution of the capability estimator \cite{oehlert1992note,van2000asymptotic,serfling2009approximation}, the resulting optimal decision rule takes the form
\[
\widehat{C}_{pk} \ge C_0 + k\,SE(\widehat{C}_{pk}),
\]
where $SE(\widehat{C}_{pk})$ denotes the estimated standard error of the capability estimator and $k$ is a calibration constant determined either by a tolerable failure probability or by a false-accept/false-reject cost ratio.

This formulation reveals that several commonly used approval procedures, such as deterministic thresholding and lower confidence bound rules, can be interpreted as special cases of a unified margin-based decision family, and further leads to a cost-sensitive approval rule derived from asymmetric operational loss. The present study focuses on the statistical decision rules governing capability-based approval under finite samples.

The main contributions of this paper are:

\begin{enumerate}
	\item We formulate capability approval as a statistical decision problem under asymmetric operational loss, explicitly distinguishing the costs of false acceptance and false rejection.
	
	\item We derive a risk-calibrated approval rule that adjusts the nominal capability threshold by a finite-sample uncertainty margin proportional to the estimated standard error of the capability estimator.
	
	\item We show that several widely used approval procedures, deterministic thresholding, lower confidence bound rules, and probability-based approval rules, can be interpreted as special cases of a unified margin-based decision framework.
	
	\item We further derive a cost-sensitive capability approval rule based on asymmetric operational loss, which provides an operational interpretation of the calibration constant in the proposed margin rule.
	
	\item Through simulation experiments, we characterize the finite-sample operating behavior of these rules and illustrate their practical implications using an industrial capability dataset.
\end{enumerate}

The remainder of the paper is organized as follows. Section~\ref{sec:preliminaries} reviews the sampling behavior of capability estimators and the boundary-instability phenomenon that motivates risk calibration. Section~\ref{sec:decision_framework} formulates capability approval as a statistical decision problem under asymmetric loss. Section~\ref{sec:unified_rules} derives the unified family of risk-calibrated approval rules and presents the main theoretical results. Section~\ref{sec:simulation} investigates the finite-sample behavior of these rules through simulation. Section~\ref{sec:case_study} presents an industrial case study illustrating the practical impact of risk-calibrated approval. Section~\ref{sec:discussion} concludes with discussion and practical guidelines. Additional technical derivations and implementation details are provided in the Appendix.

\section{Capability Estimation and Boundary Instability}
\label{sec:preliminaries}

This section reviews the sampling behavior of the capability estimator and the boundary-instability phenomenon that motivates the proposed decision framework. A detailed theoretical analysis of the boundary behavior of threshold-based capability decisions under finite samples is given in \cite{jiang2026finite}. Here we summarize the key results needed for the subsequent development.

Throughout the paper, $C_{pk}$ denotes the process capability index, $\widehat{C}_{pk}$ its estimator, and $C_{pk}^{true}$ the true process capability.

\subsection{Classical capability estimator under normal sampling}

Assume a statistically stable process with bilateral specification limits $\LSL$ and $\USL$. 
Under the classical normal-model formulation, the true process capability is
\begin{equation}
	C_{pk}^{true}
	=
	g(\mu,\sigma)
	=
	\min\left(
	\frac{\USL-\mu}{3\sigma},
	\frac{\mu-\LSL}{3\sigma}
	\right),
	\label{eq:true_cpk}
\end{equation}
where $\mu$ and $\sigma$ denote the true process mean and standard deviation. 

Given a sample $X_1,\dots,X_n$, the plug-in estimator is
\begin{equation}
	\widehat{C}_{pk}
	=
	g(\overline{X},S)
	=
	\min\left(
	\frac{\USL-\overline{X}}{3S},
	\frac{\overline{X}-\LSL}{3S}
	\right),
	\label{eq:hat_cpk}
\end{equation}

where
\[
\overline{X}=\frac{1}{n}\sum_{i=1}^n X_i,
\qquad
S^2=\frac{1}{n-1}\sum_{i=1}^n (X_i-\overline{X})^2.
\]

Under normal sampling,
\begin{equation}
	\overline{X}\sim N\!\left(\mu,\frac{\sigma^2}{n}\right),
	\qquad
	\frac{(n-1)S^2}{\sigma^2}\sim \chi^2_{n-1}.
	\label{eq:normal_sampling}
\end{equation}

Thus $\widehat{C}_{pk}$ is a random variable induced by the joint sampling variability of $\overline{X}$ and $S$.

\subsection{Local asymptotic approximation}
Under standard regularity conditions ensuring differentiability of the capability functional and joint asymptotic normality of the estimators $(\overline{X}, S)$, the capability estimator admits a local asymptotic approximation based on classical results such as the Delta Method \cite{van2000asymptotic}. This approximation is used here to characterize the sampling variability of the estimator and to facilitate the subsequent development of decision rules.

The following expansion is obtained via a standard application of the Delta Method to the capability functional:
\begin{equation}
	\sqrt{n}\bigl(\widehat{C}_{pk}-C_{pk}^{true}\bigr)
	\overset{d}{\longrightarrow}
	\mathcal{N}(0,\sigma_C^2),
	\label{eq:cpk_asymptotic_normal}
\end{equation}
where the asymptotic variance can be expressed as
\begin{equation}
	\sigma_C^2
	=
	\nabla g(\mu,\sigma)^\top \Sigma \nabla g(\mu,\sigma),
	\label{eq:sigma_c_general}
\end{equation}
with $\Sigma$ denoting the asymptotic covariance matrix of $(\overline{X},S)$. Equivalently,
\begin{equation}
	\widehat{C}_{pk}
	=
	C_{pk}^{true}
	+
	\frac{\sigma_C}{\sqrt{n}}Z
	+
	o_p(n^{-1/2}),
	\quad Z\sim \mathcal{N}(0,1).
	\label{eq:cpk_local_expansion}
\end{equation}

This expansion indicates that the estimator fluctuates around the true capability level with stochastic dispersion of order $1/\sqrt{n}$.

This representation is not introduced as a new theoretical result, but rather used as a convenient approximation that serves as an input to the subsequent decision-theoretic formulation and enables a probabilistic characterization of estimation uncertainty in capability-based decisions.

\subsection{Boundary behavior under deterministic thresholding}
The boundary behavior described here follows from standard asymptotic properties of the estimator. It is not introduced as a new theoretical phenomenon, but rather used as a motivating observation for capability approval decisions under finite-sample uncertainty.

Consider the conventional deterministic threshold rule
\begin{equation}
	D_0 = \I(\widehat{C}_{pk}\ge C_0),
	\label{eq:d0_rule}
\end{equation}
where $C_0$ denotes a fixed capability threshold.

Using the standard local asymptotic approximation introduced above, the acceptance probability satisfies
\begin{equation}
	P(\widehat{C}_{pk}\ge C_0)
	=
	\Phi\!\left(
	\frac{\sqrt{n}(C_{pk}^{true}-C_0)}{\sigma_C}
	\right)
	+o(1).
	\label{eq:accept_prob_boundary}
\end{equation}

Under the local parameterization

\begin{equation}
	C_{pk}^{true}
	=
	C_0 + \frac{h}{\sqrt{n}},
	\quad h\in\mathbb{R},
	\label{eq:local_param}
\end{equation}

the acceptance probability converges to
\begin{equation}
	P(\widehat{C}_{pk}\ge C_0)
	\to
	\Phi\!\left(\frac{h}{\sigma_C}\right).
	\label{eq:local_accept_prob}
\end{equation}

In particular, when the true capability equals the approval threshold ($C_{pk}^{true}=C_0$),
\begin{equation}
	P(\widehat{C}_{pk}\ge C_0)\to \frac12 .
	\label{eq:half_boundary}
\end{equation}

This observation highlights a practical implication for capability approval: when the true capability is close to the threshold, deterministic thresholding can induce non-negligible decision uncertainty. This motivates the need for approval rules that explicitly calibrate estimation uncertainty relative to acceptable decision risk.
\section{Decision-Theoretic Formulation of Capability Approval}
\label{sec:decision_framework}

\subsection{Decision space and states of nature}

We consider a binary action space
\[
\mathcal{A}=\{\text{Accept},\text{Reject}\}.
\]

Let $\Theta$ denote the parameter space of the true capability level.
The true state of nature is determined by whether the unknown true capability exceeds the required threshold $C_0$:
\[
\Theta_0 = \{C_{pk}^{true} < C_0\},
\qquad
\Theta_1 = \{C_{pk}^{true} \ge C_0\}.
\]

The approval problem is therefore a binary classification problem in which the unobserved state corresponds to the capability status of the process relative to the threshold.

\subsection{Standard asymmetric loss formulation}
The decision-theoretic formulation adopted here follows standard approaches in statistical decision theory. It is not intended as a novel loss construction, but rather used as a means to explicitly represent the asymmetric consequences of capability approval decisions in an interpretable and operational manner.

We define the loss function
\begin{equation}
	L(a,C_{pk}^{true})
	=
	\begin{cases}
		c_{FA}, & a=\text{Accept}\ \text{and}\ C_{pk}^{true}<C_0,\\[3pt]
		c_{FR}, & a=\text{Reject}\ \text{and}\ C_{pk}^{true}\ge C_0,\\[3pt]
		0, & \text{otherwise},
	\end{cases}
	\label{eq:loss_function}
\end{equation}
where $c_{FA}>0$ and $c_{FR}>0$ denote the costs of a false accept and a false reject, respectively.

This formulation follows a standard asymmetric (0--1 type) loss structure, where the two types of decision errors are assigned different costs.

It is often convenient to express the asymmetry through the cost ratio $\lambda = {c_{FA}}/{c_{FR}}$. Typically $\lambda>1$ in applications where false approval of an incapable process is more costly than unnecessary rejection of a capable process.

While more complex loss structures (e.g., continuous or quadratic loss functions) could be considered, we adopt this simple formulation to maintain interpretability and direct applicability in engineering contexts, where capability approval is typically implemented as a threshold-based decision.

\subsection{Posterior expected loss}
Given the above loss formulation, the decision problem can be analyzed using standard Bayesian decision theory. Let $D$ denote the observed dataset. For each action $a\in\mathcal{A}$, define the posterior expected loss (posterior risk)
\begin{equation}
	R(a\mid D)
	=
	E[L(a,C_{pk}^{true})\mid D].
	\label{eq:posterior_risk}
\end{equation}

For the action \textit{Accept},
\begin{align}
	R(\text{Accept}\mid D)
	&=
	c_{FA}P(C_{pk}^{true}<C_0\mid D).
	\label{eq:r_accept}
\end{align}

For the action \textit{Reject},
\begin{align}
	R(\text{Reject}\mid D)
	&=
	c_{FR}P(C_{pk}^{true}\ge C_0\mid D).
	\label{eq:r_reject}
\end{align}

The Bayes-optimal decision rule minimizes posterior expected loss:
\begin{equation}
	a^*(D)
	=
	\arg\min_{a\in\mathcal{A}} R(a\mid D).
	\label{eq:bayes_action}
\end{equation}

Therefore, acceptance is optimal if and only if
\begin{equation}
	R(\text{Accept}\mid D)
	\le
	R(\text{Reject}\mid D),
	\label{eq:accept_optimal_condition_1}
\end{equation}
that is,
\begin{equation}
	c_{FA}P(C_{pk}^{true}<C_0\mid D)
	\le
	c_{FR}P(C_{pk}^{true}\ge C_0\mid D).
	\label{eq:accept_optimal_condition_2}
\end{equation}

Since
\[
P(C_{pk}^{true}\ge C_0\mid D)=1-P(C_{pk}^{true}<C_0\mid D),
\]

Equation~\eqref{eq:accept_optimal_condition_2} is equivalent to
\begin{align}
	c_{FA}P(C_{pk}^{true}<C_0\mid D)
	&\le
	c_{FR}\bigl[1-P(C_{pk}^{true}<C_0\mid D)\bigr],
	\nonumber\\
	P(C_{pk}^{true}<C_0\mid D)
	&\le \frac{c_{FR}}{c_{FA}+c_{FR}}.
	\label{eq:posterior_threshold_derivation}
\end{align}

Define $\alpha = {c_{FR}}/({c_{FA}+c_{FR}}) = {1}/({1+\lambda})$. The Bayes-optimal approval rule accepts whenever
\begin{equation}
	P(C_{pk}^{true}<C_0\mid D)\le \alpha,
	\label{eq:probability_rule}
\end{equation}
and rejects otherwise.

Equation~\eqref{eq:probability_rule} yields a natural probabilistic interpretation of capability approval under the adopted loss formulation, where the process is accepted only if the posterior probability of being truly below the threshold does not exceed $\alpha$.

\subsection{Misclassification probabilities and expected operational loss}

For a fixed approval rule $\delta(D)\in\{0,1\}$, where $\delta(D)=1$ denotes acceptance, define the false-accept and false-reject probabilities at true capability level $C_{pk}^{true}$:
\begin{align}
	P_{FA}(C_{pk}^{true},n;\delta)
	&=
	P_{C_{pk}^{true}}\bigl(\delta(D)=1\bigr), \quad C_{pk}^{true}<C_0,
	\label{eq:pfa_def}
	\\
	P_{FR}(C_{pk}^{true},n;\delta)
	&=
	P_{C_{pk}^{true}}\bigl(\delta(D)=0\bigr), \quad C_{pk}^{true}\ge C_0.
	\label{eq:pfr_def}
\end{align}

These functions depend on the sampling distribution, the estimator, the sample size, and the chosen rule.

The corresponding expected operational loss is
\begin{equation}
	\begin{aligned}
		EL(C_{pk}^{true},n;\delta)
		&=
		c_{FA}\,P_{FA}(C_{pk}^{true},n;\delta)\,\I(C_{pk}^{true}<C_0) \\
		&\quad+
		c_{FR}\,P_{FR}(C_{pk}^{true},n;\delta)\,\I(C_{pk}^{true}\ge C_0).
	\end{aligned}
	\label{eq:expected_loss}
\end{equation}

Equation~\eqref{eq:expected_loss} serves as a primary performance measure in the simulation study, because it quantifies the complete trade-off between false accept and false reject under a given cost regime.

\section{Unified Risk-Calibrated Approval Rules}
\label{sec:unified_rules}

\subsection{Asymptotic posterior approximation}

To operationalize the decision rule in \eqref{eq:probability_rule}, we require an approximation to the failure probability $P(C_{pk}^{true}<C_0\mid D)$. Under regular asymptotic conditions, the estimator admits the approximate sampling distribution
\begin{equation}
	\widehat{C}_{pk} \mid C_{pk}^{true}
	\approx
	N(C_{pk}^{true},SE(\widehat{C}_{pk})^2).
	\label{eq:cpk_asymptotic_normal}
\end{equation}

Using a standard asymptotic inversion argument, the uncertainty of the true capability around the observed estimate can be approximated by
\begin{equation}
	C_{pk}^{true}\mid D
	\approx
	N\!\bigl(\widehat{C}_{pk},SE(\widehat{C}_{pk})^2\bigr).
	\label{eq:posterior_normal_approx}
\end{equation}

This approximation may be interpreted either as an approximate posterior distribution under a noninformative prior, or more generally as a confidence-distribution representation of uncertainty.

Consequently, the probability that the true capability falls below the threshold can be approximated by
\begin{align}
	P(C_{pk}^{true}<C_0\mid D)
	&\approx
	\Phi\!\left(
	\frac{C_0-\widehat{C}_{pk}}{SE(\widehat{C}_{pk})}
	\right),
	\label{eq:posterior_failure_prob}
\end{align}
where $\Phi(\cdot)$ denotes the standard normal cumulative distribution function (CDF).

Substituting \eqref{eq:posterior_failure_prob} into \eqref{eq:probability_rule}, the acceptance condition becomes
\begin{equation}
	\Phi\!\left(
	\frac{C_0-\widehat{C}_{pk}}{SE(\widehat{C}_{pk})}
	\right)
	\le \alpha .
	\label{eq:phi_rule}
\end{equation}

Applying the inverse normal CDF yields
\begin{equation}
	\frac{C_0-\widehat{C}_{pk}}{SE(\widehat{C}_{pk})}\le z_\alpha
	\quad\Longleftrightarrow\quad
	\widehat{C}_{pk}\ge C_0 - z_\alpha SE(\widehat{C}_{pk}),
	\label{eq:analytic_guardband_rule}
\end{equation}
where $z_\alpha = \Phi^{-1}(\alpha)$.

Because $\alpha<0.5$ in most practically relevant settings, we have $z_\alpha<0$, implying that the right-hand side of \eqref{eq:analytic_guardband_rule} exceeds $C_0$. Hence explicit risk control naturally introduces a positive approval margin.

\subsubsection*{Remark on distributional assumptions}
The decision-theoretic framework itself does not rely on normality of the underlying process distribution. The normal approximation above is introduced solely to obtain a closed-form approval margin and an analytically tractable rule. For non-normal processes, the probability
$P(C_{pk}^{true}<C_0\mid D)$
may instead be estimated using bootstrap methods or distribution-specific capability estimators (e.g., non-normal capability indices). In such cases the decision rule
$P(C_{pk}^{true}<C_0\mid D)\le\alpha$
remains valid, although the margin representation may be obtained numerically rather than analytically.
A simple bootstrap implementation of this procedure is summarized in Appendix~\ref{app:bootstrap}.

\subsection{Sample-size-dependent approval margin}

This representation follows directly from standard normal-based approximations and is not intended as a new theoretical result.

Define the approval margin $\Delta(n,\alpha) = (-z_\alpha)\,SE(\widehat{C}_{pk}).$ Equation~\eqref{eq:analytic_guardband_rule} can be written as $\widehat{C}_{pk} \ge C_0 + \Delta(n,\alpha).$ When $SE(\widehat{C}_{pk})\asymp \sigma_C/\sqrt{n}$, the approval margin is
\begin{equation*}
	\Delta(n,\alpha)
	\approx
	\frac{(-z_\alpha)\sigma_C}{\sqrt{n}},
\end{equation*}
showing the margin shrinks at the canonical $1/\sqrt{n}$ rate.

This approximation highlights three key properties:
\begin{enumerate}
	\item the approval margin increases as the tolerated failure probability $\alpha$ decreases;
	\item the approval margin increases with estimator dispersion $\sigma_C$;
	\item the approval margin decreases with sample size.
\end{enumerate}

Thus finite-sample risk calibration automatically produces more conservative decisions when estimation uncertainty is larger or when stronger protection against false acceptance is required.

\subsection{A unified margin family}
The preceding derivation suggests a broader family of approval rules:
\begin{equation}
	\delta_k(D)=\I\!\left(\widehat{C}_{pk}\ge C_0+k\,SE(\widehat{C}_{pk})\right),
	\quad k\in\mathbb{R}.
	\label{eq:k_family}
\end{equation}

This formulation includes several commonly used approval rules as special cases,
differing primarily in the interpretation of the calibration constant $k$:
\begin{enumerate}
	
	\item Deterministic threshold rule: $k=0$.
	
	\item Posterior probability rule: $k=-z_\alpha$, where $z_\alpha=\Phi^{-1}(\alpha)$ is the standard normal quantile associated with the tolerated failure probability $\alpha$.
	
	\item Cost-sensitive rule: when asymmetric loss is specified through the cost ratio
	$\lambda=c_{FA}/c_{FR}$, the corresponding tolerated failure probability is
	$\alpha=1/(1+\lambda)$, yielding $k=-z_{1/(1+\lambda)}$, where $z_\alpha=\Phi^{-1}(\alpha)$ denotes the standard normal quantile and the mapping from $\lambda$ to $k$ is induced through the associated risk level $\alpha$.
	
	\item Lower confidence bound rule: $k=z_{1-\gamma}$, corresponding to approval based on the one-sided lower confidence bound $LCB_{1-\gamma}$.
	
\end{enumerate}

Thus, these approval procedures share a common functional form and differ primarily in the interpretation and calibration of the constant $k$.

\paragraph{Equivalent margin representation.}

This representation follows directly from the normal approximation in \eqref{eq:cpk_asymptotic_normal} and the failure probability approximation in \eqref{eq:posterior_failure_prob}, and is not intended as a new theoretical result.

The probability-based approval rule in \eqref{eq:probability_rule} can be written in the following equivalent margin form:
\[
\widehat{C}_{pk} \ge C_0 + k_\alpha \, SE(\widehat{C}_{pk}),
\quad
k_\alpha = -z_\alpha,
\]
where $z_\alpha=\Phi^{-1}(\alpha)$ denotes the standard normal quantile associated with the risk level $\alpha$.

This form follows directly from \eqref{eq:phi_rule}--\eqref{eq:analytic_guardband_rule} and the posterior threshold derivation in \eqref{eq:posterior_threshold_derivation}.

Under an asymmetric loss ratio $\lambda = c_{FA}/c_{FR}$, the corresponding decision rule is obtained by setting
\[
\alpha = \frac{1}{1+\lambda},
\quad
k_\lambda = -z_{1/(1+\lambda)}.
\]

This representation highlights that commonly used approval rules, including deterministic thresholding, confidence-bound rules, and probability-based rules, can be interpreted within a unified margin-based structure, providing a common perspective on their calibration and interpretation.

\subsection{Local asymptotic interpretation of the calibration constant}

The family \eqref{eq:k_family} admits a useful interpretation under a local asymptotic regime. Suppose the capability estimator admits the local asymptotic expansion in \cite{jiang2026finite}:
\begin{equation}
	\widehat{C}_{pk}
	=
	C_{pk}^{true}+\frac{\sigma_C}{\sqrt{n}}Z+o_p(n^{-1/2}),
	\quad Z\sim \mathcal{N}(0,1),
	\label{eq:local_for_k}
\end{equation}
and $SE(\widehat{C}_{pk})\approx\sigma_C/\sqrt{n}$.  Then under the rule $\delta_k(D)$ defined in \eqref{eq:k_family},
\begin{align}
	P(\delta_k(D)=1)
	&=
	P\!\left(\widehat{C}_{pk}\ge C_0+k\,SE(\widehat{C}_{pk})\right)
	\nonumber\\
	&\approx
	\Phi\!\left(
	\frac{\sqrt{n}(C_{pk}^{true}-C_0)}{\sigma_C}-k
	\right).
	\label{eq:accept_prob_k}
\end{align}

Now consider the local parameterization
\[
C_{pk}^{true}=C_0+\frac{h}{\sqrt{n}}, \quad h \in \mathbb{R}.
\]
Then
\begin{equation}
	P(\delta_k(D)=1)
	\to
	\Phi\!\left(\frac{h}{\sigma_C}-k\right).
	\label{eq:local_accept_prob_k}
\end{equation}

In particular, at the boundary $C_{pk}^{true}=C_0$,
\begin{equation}
	P(\delta_k(D)=1)\to\Phi(-k).
	\label{eq:boundary_accept_prob_k}
\end{equation}

Thus the calibration constant $k$ directly determines the asymptotic boundary acceptance probability. 

For example:
\begin{itemize}
	\item $k=0$ gives $\Phi(0)=0.5$;
	\item $k=1.645$ gives $\Phi(-1.645)\approx0.05$;
	\item $k=2.326$ gives $\Phi(-2.326)\approx0.01$.
\end{itemize}

This yields the following result.

\begin{theorem}[Boundary Risk Calibration]
	\label{thm:boundary_calibration}
	Suppose \eqref{eq:local_for_k} holds and $SE(\widehat{C}_{pk})\approx\sigma_C/\sqrt{n}$. Consider the rule
	\[
	\delta_k(D)=\I\!\left(\widehat{C}_{pk}\ge C_0+k\,SE(\widehat{C}_{pk})\right).
	\]
	Then
	\[
	P(\delta_k(D)=1)
	=
	\Phi\!\left(
	\frac{\sqrt{n}(C_{pk}^{true}-C_0)}{\sigma_C}-k
	\right)
	+o(1),
	\quad n \rightarrow \infty
	\]
	In particular, if $C_{pk}^{true}=C_0$, then
	\[
	P(\delta_k(D)=1)\to\Phi(-k).
	\]
	Choosing
	\[
	k=z_{1-\alpha}
	\]
	ensures that the asymptotic boundary acceptance probability equals $\alpha$.
\end{theorem}

\subsection{Connection to lower confidence bounds}
The margin representation also provides a natural interpretation of classical lower confidence bound approval rules. Consider the one-sided lower confidence bound
\begin{equation}
	LCB_{1-\gamma}
	=
	\widehat{C}_{pk}-z_{1-\gamma}SE(\widehat{C}_{pk}).
	\label{eq:lcb_def}
\end{equation}

The rule $LCB_{1-\gamma}\ge C_0$ is equivalent to
\begin{equation}
	\widehat{C}_{pk}\ge C_0+z_{1-\gamma}SE(\widehat{C}_{pk}).
\end{equation}

Hence the LCB rule is a member of the family \eqref{eq:k_family} with $k_{LCB}=z_{1-\gamma}$. Thus LCB methods, probability-based rules, and cost-sensitive rules all modify the deterministic threshold through a common guard-band mechanism. The difference lies primarily in the interpretation of the calibration constant:
\begin{itemize}
	\item inferential coverage for LCB rules,
	\item posterior failure risk for probability rules,
	\item operational cost asymmetry for cost-sensitive rules.
\end{itemize}

\subsection{Practical selection of $\lambda$, $\alpha$, and $k$}

Since
\[
\alpha=\frac{1}{1+\lambda},
\qquad
k=-z_\alpha=z_{1-\alpha},
\]
the operational loss ratio determines the guard-band constant through a one-to-one mapping between $\lambda$, $\alpha$, and $k$. Typical values of these quantities are summarized in Table~\ref{tab:lambda_alpha_k}. The numerical values of $k$ reported here correspond to standard normal quantiles and are included for illustration only; they do not constitute new statistical results.
\begin{table}[h]
	\centering
	\caption{Relationship between the operational loss ratio $\lambda$, the tolerated failure probability $\alpha$, and the corresponding calibration constant $k$, based on standard normal quantiles.}
	\label{tab:lambda_alpha_k}
	\begin{tabular*}{\columnwidth}{@{\extracolsep{\fill}}ccc}
		\toprule
		Loss ratio ($\lambda$) & Failure prob. ($\alpha$) & Calib. constant ($k$) \\
		\midrule
		1  & 0.50 & 0 \\
		4  & 0.20 & 0.842 \\
		9  & 0.10 & 1.282 \\
		19 & 0.05 & 1.645 \\
		99 & 0.01 & 2.326 \\
		\bottomrule
	\end{tabular*}
\end{table}

These values illustrate how standard risk levels map to different approval margins under the normal approximation.

\subsection{Practical elicitation of cost ratio}

The operational loss ratio $\lambda = c_{FA}/c_{FR}$ determines the level of protection against false acceptance. In practice, it reflects the relative importance of external quality risk versus internal efficiency loss.

False acceptance (approving an incapable process) may lead to customer complaints, warranty claims, or safety risks, whereas false rejection typically incurs internal costs such as rework, additional inspection, or production delays. Rather than requiring precise monetary valuation, $\lambda$ serves as a relative measure of these competing consequences.

Typical ranges of $\lambda$ depend on the application context:

\paragraph{Safety-critical components.}
For automotive, aerospace, or medical applications, false acceptance can have severe consequences. In such cases, $\lambda$ is typically large (e.g., $\lambda \ge 20$), leading to conservative approval rules.

\paragraph{High-volume consumer products.}
For consumer products, false acceptance leads to warranty-related costs, while false rejection affects yield and efficiency. Moderate values (e.g., $\lambda$ between 5 and 10) are often appropriate.

\paragraph{Non-critical features.}
For cosmetic or non-functional dimensions, the cost of false acceptance is relatively low. In such cases, $\lambda$ may be close to 1, resulting in approval rules similar to deterministic thresholding.

These examples highlight that $\lambda$ is context-dependent. The proposed framework provides a transparent mechanism for incorporating such considerations into capability approval decisions.

\section{Finite-Sample Simulation Study}
\label{sec:simulation}

This section investigates the finite-sample operating characteristics of the proposed approval rules and shows how risk calibration alters approval behavior, error trade-offs, and expected operational loss under repeated sampling. The simulation study serves two purposes. First, it quantifies how deterministic threshold, lower confidence bound, and risk-calibrated rules behave when capability is estimated from finite samples. Second, it examines how asymmetric operational costs reshape the false-accept/false-reject trade-off and the resulting expected operational loss.

The simulation addresses three questions:
\begin{enumerate}
	\item How do deterministic threshold, lower confidence bound, and risk-calibrated approval rules differ under repeated sampling?
	\item How do false-accept and false-reject probabilities vary with the true capability level, sample size, and cost ratio?
	\item Under which operating regimes does risk calibration reduce expected operational loss relative to deterministic thresholding?
\end{enumerate}

\subsection{Simulation setup}

\subsubsection{Baseline data-generating model}

In the main simulation, samples are generated under the normal model with bilateral specifications. To isolate the effect of estimator variability and obtain a transparent mapping between the target capability level and the process variance, we use symmetric specification limits
\[
\LSL=-T,\qquad \USL=T,
\]
together with a centered process mean $\mu=0.$ Under this construction, the true process capability reduces to $C_{pk}^{true} ={T}/{3\sigma}.$ For a target capability level $C_{pk}^{true}$, the corresponding process standard deviation is therefore chosen as $\sigma = {T}/{3C_{pk}^{true}}.$ Thus each target capability level corresponds to the normal sampling model
\[
X_i \sim \mathcal{N}\!\left(0,\left(\frac{T}{3C_{pk}^{true}}\right)^2\right).
\]

Under this centered symmetric construction, the capability index is determined by a single active constraint (either $\USL$ or $\LSL$), which is consistent with the conditions under which the closed-form variance expression and the corresponding standard error approximation are derived. This corresponds to the regular regime assumed in the asymptotic variance derivation.

This setting is adopted to obtain an analytically tractable and transparent relationship between the target capability level and the sampling distribution of the estimator, allowing the effect of estimation uncertainty to be studied in a controlled manner.

In the numerical implementation, we set $T=4$. In this setting, the specific value of $T$ is immaterial because the capability index is scale-free with respect to $T$ under the chosen normalization.

\subsubsection{Parameter grid}

The simulation spans a range of practically relevant configurations:
\[
C_{pk}^{true}\in\{0.80,0.82,\dots,2.00\}, n\in\{20,32,50,80,120,200\},
\]
with approval threshold $C_0=1.33.$ The fine grid in $C_{pk}^{true}$ is chosen to resolve the transition region around the approval boundary with sufficient precision. The value $n=32$ is of particular practical interest because it matches a common engineering audit size and is also used in the industrial case study.

To study asymmetric operational loss, we consider
\[
\lambda\in\{1,2,5,10,20,50,100\},
\]
which represent progressively stronger penalties for false acceptance. Monte Carlo estimates are based on $B=10{,}000$ replications per configuration, unless otherwise stated.

\subsubsection{Capability estimation and standard error approximation}

For each configuration $(C_{pk}^{\mathrm{true}}, n)$, an independent sample
$X_1,\dots,X_n$ is generated from the calibrated normal model. The estimator
$\widehat{C}_{pk}$ defined in \eqref{eq:hat_cpk} is then computed using the
sample mean $\overline{X}$ and sample standard deviation $S$.

The decision rules introduced in Section~\ref{sec:unified_rules} also require an estimate of the standard error of $\widehat{C}_{pk}$. In the main simulation, we use the analytic plug-in approximation $SE(\widehat{C}_{pk})\approx {\widehat{\sigma}_C}/{\sqrt{n}}$. Following \cite{jiang2026finite}, under the regular one-sided active-constraint regime the asymptotic variance of the capability estimator is
\[
\sigma_C^2
=
\frac19 + \frac{(C_{pk}^{true})^2}{2}.
\]
The derivation of this expression is provided in Appendix~\ref{app:sigma}. The variance expression is derived under a centered process with symmetric specifications ($\mu = 0$) and a locally one-sided active-constraint regime, under which the capability functional is differentiable and admits a closed-form asymptotic variance. When the process mean deviates from the target ($\mu \neq 0$), both constraints may become locally active, and the approximation may become less accurate, potentially introducing bias in the estimated standard error; in such cases, more general variance expressions or bootstrap-based estimation can be used.

Using a plug-in estimator, the standard error of $\widehat{C}_{pk}$ is approximated by
\[
SE(\widehat{C}_{pk})
\approx
\sqrt{\frac{1/9+\widehat{C}_{pk}^2/2}{n}}.
\]

This approximation is computationally efficient and directly aligned with the asymptotic theory developed earlier. A bootstrap-based standard error was also examined as a robustness check; details are discussed in the non-normal extension below.

\subsubsection{Decision rules compared}

All rules are evaluated using the same simulated samples and the same estimator $\widehat{C}_{pk}$, differing only in the acceptance criterion.

\paragraph{Rule 1: Deterministic threshold rule.}
\begin{equation}
	\delta_{det}
	=
	\I(\widehat{C}_{pk}\ge C_0).
	\label{eq:sim_det}
\end{equation}

\paragraph{Rule 2: Lower confidence bound rule.}
For a one-sided confidence level $1-\gamma$, define
\begin{equation}
	LCB_{1-\gamma}
	=
	\widehat{C}_{pk} - z_{1-\gamma}SE(\widehat{C}_{pk}),
	\label{eq:sim_lcb}
\end{equation}
and accept if
\begin{equation}
	\delta_{LCB,\gamma}
	=
	\I(LCB_{1-\gamma}\ge C_0).
	\label{eq:sim_lcb_rule}
\end{equation}
Unless otherwise noted, the illustrative LCB rule uses $\gamma=0.05$.

\paragraph{Rule 3: Probability rule.}
For a target failure probability $\alpha$, define
\begin{equation}
	p_{fail}
	=
	\Phi\!\left(\frac{C_0-\widehat{C}_{pk}}{SE(\widehat{C}_{pk})}\right),
	\label{eq:sim_pfail}
\end{equation}
and accept if
\begin{equation}
	\delta_{prob,\alpha}
	=
	\I(p_{fail}\le \alpha).
	\label{eq:sim_prob_rule}
\end{equation}

\paragraph{Rule 4: Cost-sensitive rule.}
For cost ratio $\lambda=c_{FA}/c_{FR}$, define
\[
\alpha=\frac{1}{1+\lambda},
\qquad
k_\lambda=-z_\alpha.
\]
Then accept if
\begin{equation}
	\delta_{\lambda}
	=
	\I\!\left(
	\widehat{C}_{pk}\ge C_0 + k_\lambda SE(\widehat{C}_{pk})
	\right).
	\label{eq:sim_cost_rule}
\end{equation}

Under the normal approximation, the probability rule and the cost-sensitive rule are analytically equivalent after the mapping $\alpha=1/(1+\lambda)$. In the results below, we emphasize the cost-sensitive parameterization because it is more directly interpretable in operational terms.

\subsubsection{Performance metrics}

For each configuration $(C_{pk}^{true},n)$ and approval rule $\delta$, we record the acceptance probability
\begin{equation*}
	P_{acc}(C_{pk}^{true},n;\delta)
	\approx
	\frac{1}{B}\sum_{b=1}^B \delta^{(b)}.
\end{equation*}

When $C_{pk}^{true}<C_0$, acceptance corresponds to a false accept, so the false-accept probability is estimated by
\begin{equation*}
	\widehat{P}_{FA}(C_{pk}^{true},n;\delta)
	=
	\frac{1}{B}\sum_{b=1}^B \delta^{(b)}.
\end{equation*}
When $C_{pk}^{true}\ge C_0$, rejection corresponds to a false reject, so the false-reject probability is estimated by
\begin{equation*}
	\widehat{P}_{FR}(C_{pk}^{true},n;\delta)
	=
	\frac{1}{B}\sum_{b=1}^B (1-\delta^{(b)}).
\end{equation*}

Because acceptance probability alone does not distinguish correct from incorrect decisions, the primary comparative metrics are $\widehat{P}_{FA}$, $\widehat{P}_{FR}$, and the expected operational loss
\begin{equation}
	\begin{aligned}
		\widehat{EL}(C_{pk}^{\mathrm{true}},n;\delta)
		&=
		c_{FA}\widehat{P}_{FA}(C_{pk}^{\mathrm{true}},n;\delta)
		\I(C_{pk}^{\mathrm{true}}<C_0) \\
		&\quad+
		c_{FR}\widehat{P}_{FR}(C_{pk}^{\mathrm{true}},n;\delta)
		\I(C_{pk}^{\mathrm{true}}\ge C_0).
	\end{aligned}
	\label{eq:sim_el}
\end{equation}

Using this setup, we examine three complementary aspects of finite-sample performance: the geometry of the acceptance region, the induced false-accept/false-reject trade-off, and the resulting expected operational loss.

\subsection{Acceptance probability and effective approval boundaries}

\begin{figure*}[t]
	\centering
	\includegraphics[width=0.9\textwidth]{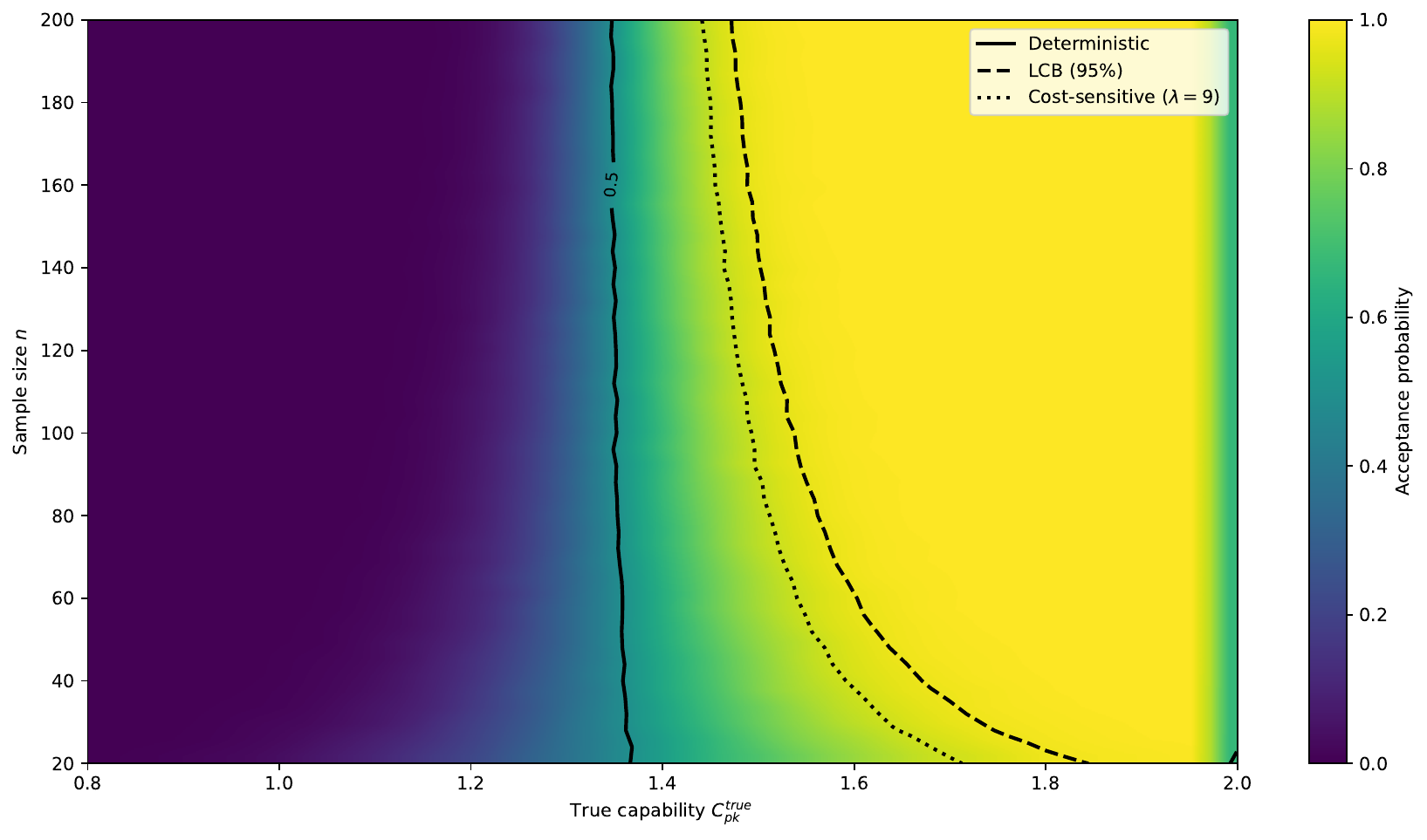}
	\caption{
		Acceptance probability surface under deterministic threshold approval (background heatmap).
		The deterministic threshold rule accepts when $\widehat{C}_{pk} \ge C_0$.
		The color scale represents the probability of approval under repeated sampling.
		Overlaid contours show the effective approval boundaries defined by 
		$P(\mathrm{Accept}) = 0.5$ for three rules:
		the deterministic threshold rule (solid),
		the $95\%$ lower confidence bound rule (dashed),
		and the cost-sensitive rule with $\lambda = 9$ (dotted).
		Results are based on Monte Carlo simulation under normal sampling with
		symmetric specification limits and centered processes.
	}
	\label{fig:acceptance_surface}
\end{figure*}

Figure~\ref{fig:acceptance_surface} displays the acceptance behavior of the approval rules across a range of true capability levels and sample sizes. The background heatmap corresponds to deterministic thresholding. Consistent with the boundary-instability result developed earlier, the $0.5$ acceptance contour under deterministic threshold approval lies close to the nominal threshold $C_{pk}^{true}=C_0$, while the transition region narrows as $n$ increases.

Two patterns are especially clear. First, the transition from low to high acceptance probability is not abrupt, but spread over a nontrivial neighborhood of the threshold, particularly when the sample size is small. This reflects the sampling variability of $\widehat{C}_{pk}$ and confirms that threshold-based approval remains intrinsically probabilistic in moderate samples. Second, both the LCB and cost-sensitive rules shift the effective approval boundary to the right. Equivalently, once uncertainty is incorporated explicitly, a larger true capability level is required to attain the same approval probability.

The magnitude of this rightward shift is greatest for small sample sizes, where estimator dispersion is largest. Figure~\ref{fig:acceptance_surface} therefore provides a geometric visualization of the unified margin family $\widehat{C}_{pk} \ge C_0 + k\,SE(\widehat{C}_{pk}),$ showing that larger calibration constants $k$ induce more conservative finite-sample approval boundaries.

\subsection{False-accept/false-reject trade-offs under asymmetric cost}

\begin{figure*}[t]
	\centering
	\subfloat[False-accept probability $P_{FA}$ as a function of the cost ratio $\lambda$]{
		\label{fig:pfa_lambda}
		\includegraphics[width=0.475\textwidth]{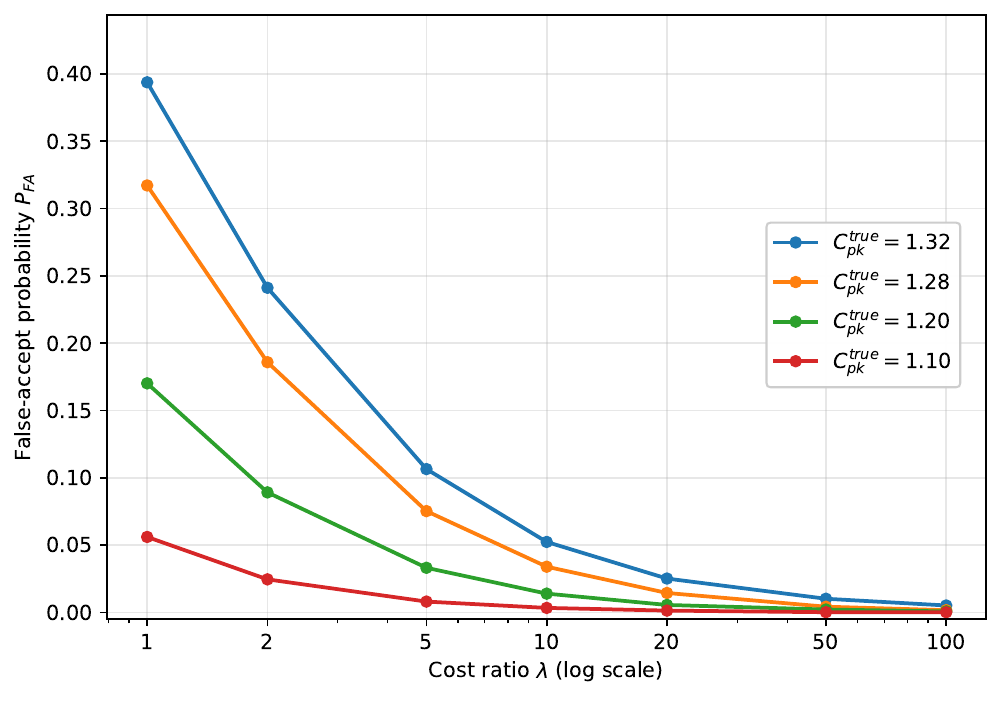}
	}
	\hspace{2pt}
	\subfloat[False-reject probability $P_{FR}$ as a function of the cost ratio $\lambda$]{
		\label{fig:pfr_lambda}
		\includegraphics[width=0.475\textwidth]{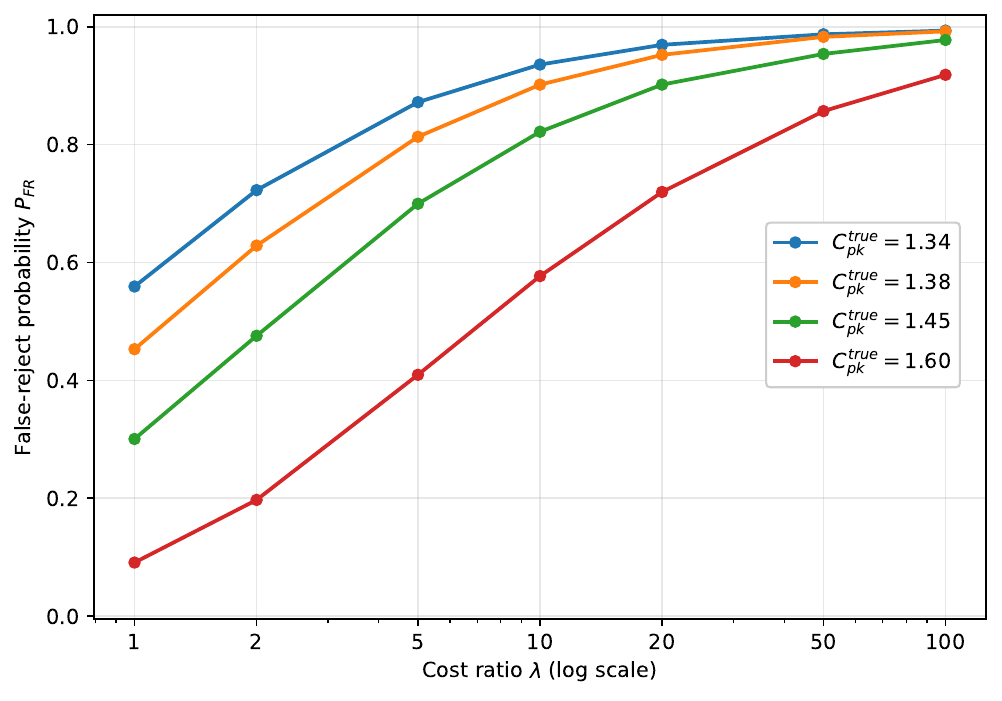}
	}
	\caption{
		Error probabilities under the cost-sensitive approval rule as the cost ratio $\lambda = c_{FA}/c_{FR}$ varies.
		The horizontal axis is shown on a logarithmic scale.
		Results are based on Monte Carlo simulation under normal sampling with threshold $C_0 = 1.33$, sample size $n = 32$, and replication size $B = 10{,}000$.
		(a) False-accept probability for selected sub-threshold capability levels ($C_{pk}^{true} < C_0$).
		(b) False-reject probability for selected supra-threshold capability levels ($C_{pk}^{true} \ge C_0$).
		As $\lambda$ increases, the acceptance region contracts, reducing false acceptance at the cost of more frequent false rejection.
	}
	\label{fig:error_lambda_tradeoff}
\end{figure*}

Figure~\ref{fig:error_lambda_tradeoff} quantifies the error trade-off induced by cost-sensitive approval as the cost ratio $\lambda$ varies. Larger values of $\lambda$ correspond to stronger penalties for false acceptance and therefore to more conservative approval rules.

Figure~\ref{fig:pfa_lambda} shows the false-accept probability for several sub-threshold processes. As $\lambda$ increases, the acceptance region contracts monotonically, producing a rapid decline in $P_{FA}$ across all sub-threshold capability levels. The reduction is most pronounced for processes close to the boundary, such as $C_{pk}^{true}=1.32$, where sampling variability would otherwise produce frequent erroneous approvals under less conservative rules.

Figure~\ref{fig:pfr_lambda} shows the corresponding false-reject probability for several supra-threshold processes. As expected, the same increase in conservatism that lowers $P_{FA}$ also raises $P_{FR}$. This increase is again largest for processes near the threshold, where modest changes in the approval margin materially affect the decision outcome.

Taken together, these results illustrate the fundamental trade-off created by asymmetric operational costs. Strengthening protection against false acceptance necessarily increases the probability of rejecting truly capable processes. The cost ratio $\lambda$ therefore acts as a transparent calibration parameter that translates operational priorities into a statistically explicit decision rule.

\subsection{Expected-loss comparison}

\begin{figure*}[!t]
	\centering
	\includegraphics[width=0.9\textwidth]{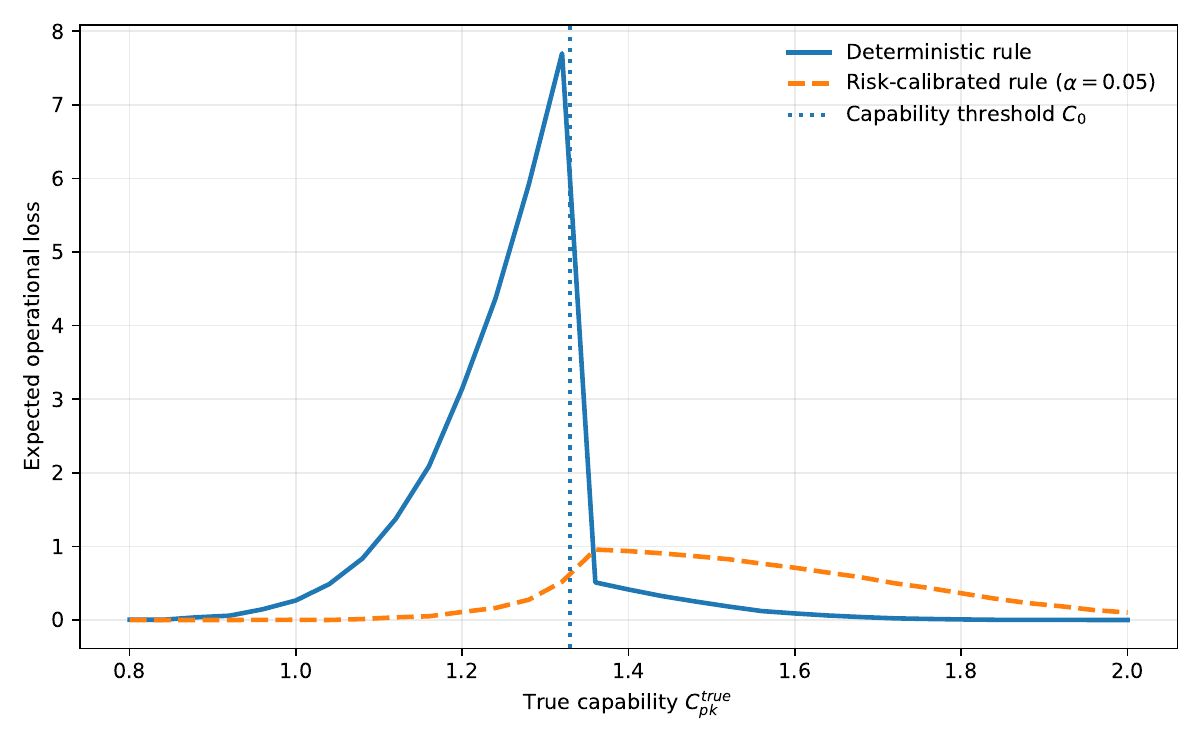}
	\caption{
		Expected operational loss under deterministic threshold and risk-calibrated approval rules.
		The deterministic threshold rule accepts when $\widehat{C}_{pk} \ge C_0$, whereas the risk-calibrated rule corresponds to the probability rule with $\alpha = 0.05$.
		Under this calibration, the probability rule, the $95\%$ lower confidence bound rule, and the cost-sensitive rule with $\lambda = 19$ are equivalent.
		Results are based on Monte Carlo simulation with sample size $n = 32$ and $B = 12{,}000$ replications under normal sampling.
		The vertical dashed line indicates the approval threshold $C_0 = 1.33$.
	}
	\label{fig:expected_loss}
\end{figure*}

Figure~\ref{fig:expected_loss} compares the expected operational loss of deterministic threshold approval and a risk-calibrated rule under the cost regime corresponding to $\alpha=0.05$, equivalently $\lambda=19$. This figure aggregates the false-accept/false-reject trade-off into a single decision-theoretic performance metric.

For quantitative reference, the corresponding expected loss values for selected cost ratios are reported in Table~\ref{tab:expected_loss_values}, where the reduction in expected loss exceeds 90\% for $\lambda \ge 20$.
\begin{table}[h]
	\centering
	\caption{Expected operational loss under deterministic and risk-calibrated approval rules for selected cost ratios ($n=32$, $C_{pk}^{true}=1.33$). Results are based on Monte Carlo simulation with $B=12{,}000$ replications.}
	\label{tab:expected_loss_values}
	\begin{tabular*}{\columnwidth}{@{\extracolsep{\fill}}cccc}
		\toprule
		$\lambda$ & Deterministic & Risk-calibrated & Reduction (\%) \\
		\midrule
		2   & 0.793 & 0.483 & 39.16 \\
		5   & 2.037 & 0.565 & 72.29 \\
		10  & 4.021 & 0.553 & 86.26 \\
		20  & 8.052 & 0.527 & 93.46 \\
		50  & 20.038 & 0.479 & 97.61 \\
		100 & 40.567 & 0.375 & 99.08 \\
		\bottomrule
	\end{tabular*}
\end{table}

This table complements Figure~\ref{fig:expected_loss} by providing explicit numerical values of expected loss across cost ratios.

The largest performance differences occur in a narrow neighborhood of the approval threshold, where sampling variability and decision asymmetry interact most strongly. In the sub-threshold region, deterministic thresholding incurs substantial loss because marginally incapable processes may still be accepted with non-negligible probability. When false acceptance is heavily penalized, these erroneous approvals dominate the expected-loss profile.

By contrast, the risk-calibrated rule substantially reduces expected loss below the threshold by introducing an explicit approval margin that accounts for estimation uncertainty. This protection is achieved at the cost of somewhat greater conservatism just above the threshold, where expected loss may increase slightly because of additional false rejections. Nevertheless, near the boundary---the regime of greatest practical concern---the overall expected loss remains markedly lower under risk calibration.

These results show that the proposed framework does not merely redistribute error probabilities. Under asymmetric operational cost, it can improve overall decision performance relative to conventional deterministic thresholding.

\subsection{Robustness to non-normality: simulation study}

To assess robustness, we extend the simulation study to non-normal data-generating mechanisms and evaluate the performance of the proposed approval rules under such settings. The simulation framework is adapted by replacing the normal-based capability index with a capability quantity appropriate to the underlying model. A practical first extension is shifted lognormal sampling:
\[
Y_i\sim \log N(m,s^2), \qquad X_i = Y_i - c,
\]
with parameters calibrated so that the percentile-based capability index equals a target value $C_{Npk}^{true}$.

Under non-normality, two approaches are especially relevant:
\begin{enumerate}
	\item an analytic rule based on an approximate standard error;
	\item a bootstrap probability rule based on the empirical failure probability.
\end{enumerate}

For the bootstrap implementation, each synthetic dataset is resampled $B_{boot}$ times, yielding capability estimates $\widehat{C}_{pk}^{*(1)},\dots,\widehat{C}_{pk}^{*(B_{boot})}$. The empirical failure probability is then estimated by
\[
\widehat{p}_{fail}
=
\frac{1}{B_{boot}}\sum_{b=1}^{B_{boot}} I(\widehat{C}_{pk}^{*(b)}<C_0),
\]
and approval is granted when $\widehat{p}_{fail}\le \alpha.$

In the simulation, we compare the deterministic threshold rule, the analytic probability rule, and the bootstrap-based rule under the same cost ratio, evaluating performance in terms of empirical acceptance behavior and expected operational loss. The results indicate that the analytic rule remains reasonably well calibrated under mild departures from normality, although noticeable deviations arise for highly skewed distributions. By contrast, the bootstrap-based rule provides more accurate control of the failure probability in the simulated settings, as it does not rely on the normal approximation.

Overall, risk-calibrated approval consistently reduces expected operational loss relative to deterministic thresholding, particularly in near-threshold regimes and under asymmetric cost.

\begin{table}[h]
	\centering
	\small
	\caption{Performance comparison of capability approval rules under lognormal sampling ($n=32$, $\lambda=10$). Results are based on Monte Carlo simulation with $B=3000$ replications. The table reports empirical acceptance-related metrics and expected operational loss ($EL$).}
	\label{tab:non_normal_performance}
	\begin{tabular*}{\columnwidth}{@{\extracolsep{\fill}}lccc}
		\toprule
		Method & Acceptance prob. & Rejection prob. & $EL$ \\
		\midrule
		Deterministic & 0.9903 & 0.0097 & 9.913 \\
		Analytic rule & 0.9447 & 0.0553 & 9.502 \\
		Bootstrap rule & 0.9127 & 0.0873 & 9.214 \\
		\bottomrule
	\end{tabular*}
\end{table}

Table~\ref{tab:non_normal_performance} shows that both analytic and bootstrap-based risk-calibrated rules reduce expected operational loss relative to deterministic thresholding under non-normal sampling, with the bootstrap rule providing the most consistent improvement.

\section{Industrial Case Study}
\label{sec:case_study}

To illustrate the practical implications of the proposed framework, we apply the risk-calibrated approval methodology to an industrial manufacturing dataset with 880 dimensions. All measurements, nominal values, and specification limits are shifted by a constant offset, which does not affect the capability analysis. The objective is to examine how approval outcomes change when deterministic thresholding is replaced by uncertainty-adjusted, cost-sensitive decision rules.

The case study addresses three empirical questions:
\begin{enumerate}
	\item How many dimensions change approval status when deterministic thresholding is replaced by risk-calibrated approval?
	\item Are reclassifications concentrated near the capability threshold, where sampling uncertainty is most influential?
	\item Does risk-calibrated approval improve empirical decision stability and reduce estimated operational loss under asymmetric cost assumptions?
\end{enumerate}

\subsection{Dataset and empirical setting}

The dataset contains measurements for $J$ product dimensions collected from a manufacturing process. For each dimension $j=1,\dots,J$, the dataset provides
\begin{itemize}
	\item a sample of $n_j$ observations,
	\item bilateral specification limits $\LSL_j$ and $\USL_j$,
	\item an estimated process capability index based on the observed sample.
\end{itemize}

In the present application, the dataset contains $J = 880$ dimensions. For each dimension, a sample of $n_j = 32$ observations is available together with the corresponding specification limits. The nominal approval threshold is $C_0 = 1.33,$ which is a commonly used capability requirement in industrial quality control. 

\begin{table}[t]
	\caption{
		Summary statistics of the industrial capability dataset used in the empirical study.
		The dataset contains measurements from $880$ product dimensions, each with
		$n=32$ observations. Dimensions were classified as approximately normal
		($J_N=582$) or non-normal ($J_{NN}=298$) based on a normality test.
		For each dimension, the capability index $\widehat{C}_{pk}$ was estimated
		using the sample mean and standard deviation.
		The approval threshold used in the analysis is $C_0 = 1.33$.
	}
	\label{tab:dataset_summary}
	\centering
	\begin{tabular*}{\columnwidth}{@{\extracolsep{\fill}}lcc}
		\hline
		Categories & Normal & Non-normal \\
		\hline
		
		\multicolumn{3}{l}{\textit{Dataset composition}} \\
		Number of dimensions & 582 & 298 \\
		Sample size per dimension ($n_j$) & 32 & 32 \\
		Capability threshold ($C_0$) & \multicolumn{2}{c}{1.33} \\
		
		\hline
		\multicolumn{3}{l}{\textit{Capability statistics}} \\
		Mean estimated capability $\overline{\widehat{C}}_{pk}$ & 2.367 & 1.848 \\
		Median estimated capability & 1.325 & 0.758 \\
		Standard deviation of $\widehat{C}_{pk}$ & 3.116 & 3.862 \\
		Minimum estimated capability & -4.971 & -6.790 \\
		Maximum estimated capability & 15.881 & 34.228 \\
		
		\hline
		\multicolumn{3}{l}{\textit{Threshold-related statistics}} \\
		Below $C_0$ ($\widehat{C}_{pk}<C_0$) & 0.500 & 0.648 \\
		Near $C_0$ ($|\widehat{C}_{pk}-C_0|\le0.05$) & 0.021 & 0.023 \\
		Near $C_0$ ($|\widehat{C}_{pk}-C_0|\le0.10$) & 0.043 & 0.054 \\
		Near $C_0$ ($|\widehat{C}_{pk}-C_0|\le0.15$) & 0.055 & 0.067 \\
		Near $C_0$ ($|\widehat{C}_{pk}-C_0|\le0.20$) & 0.076 & 0.084 \\
		\hline
	\end{tabular*}
\end{table}
Table~\ref{tab:dataset_summary} summarizes the capability characteristics of the industrial dataset. Among the 880 dimensions analyzed, $J_N=582$ were classified as approximately normal and $J_{NN}=298$ as non-normal based on a normality test.

The median capability of the normal subset is $1.325$, which lies very close to the approval threshold $C_0=1.33$. This indicates that many dimensions operate near the decision boundary, making the dataset particularly suitable for examining finite-sample approval behavior. The threshold-band statistics reported in Table~\ref{tab:dataset_summary} show that a meaningful subset of dimensions operates close to the approval threshold. Such near-threshold observations are particularly important because, as shown in Section~\ref{sec:preliminaries}, deterministic capability decisions are most sensitive to sampling variability in this regime. 

Although the overall proportion of such dimensions is relatively small, reclassifications under risk-calibrated approval are concentrated almost entirely within this narrow region around $C_0$, indicating that these cases play a disproportionate role in practical decision outcomes.

Very large capability estimates occasionally occur when the observed process variability is extremely small relative to the specification width. In industrial capability analysis, such cases are often examined together with measurement system resolution and gauge capability. In the present study, these values are retained because the focus is on comparing approval rules rather than on diagnosing measurement systems, and because they do not affect the near-threshold decision behavior of primary interest.

For consistency with the theoretical framework, capability estimates for all dimensions are computed using the plug-in estimator in \eqref{eq:hat_cpk}, ensuring that differences in approval outcomes arise from the decision rule rather than alternative capability definitions. For approximately normal dimensions, failure probabilities are estimated using the analytic normal approximation in Section~\ref{sec:unified_rules}. For non-normal dimensions, the sampling distribution of $\widehat{C}_{pk}$ may deviate from the analytic approximation. In those cases, failure probabilities are estimated using bootstrap resampling, which avoids reliance on distributional assumptions while maintaining a common capability-based decision framework across all dimensions.

\subsection{Dimension-wise capability estimation}

For each dimension $j$, the capability index is estimated using the standard plug-in estimator
\begin{equation}
	\widehat{C}_{pk, j}
	=
	\min\left(
	\frac{\USL_j-\overline{X}_j}{3S_j},
	\frac{\overline{X}_j-\LSL_j}{3S_j}
	\right),
	\label{eq:dimension_capability}
\end{equation}
where $\overline{X}_j$ and $S_j$ denote the sample mean and sample standard deviation of the $j$-th dimension.

To quantify estimation uncertainty, we compute an estimated failure probability for each dimension. For dimensions classified as approximately normal, we use the analytic approximation
\begin{equation}
	\widehat{p}_{fail,j}
	=
	\Phi\!\left(
	\frac{C_0-\widehat{C}_{pk, j}}{SE(\widehat{C}_{pk,j})}
	\right),
	\label{eq:dimension_fail_prob}
\end{equation}
which approximates the probability that the true capability level lies below the approval threshold. For dimensions classified as non-normal, the corresponding failure probability is estimated by bootstrap resampling. In both cases, the resulting quantity provides a direct measure of decision risk for each dimension.

\subsection{Reclassification under risk-calibrated approval}

Under the conventional deterministic threshold rule, dimension $j$ is accepted whenever $\widehat{C}_{pk, j} \ge C_0.$

Under the cost-sensitive risk-calibrated rule, approval depends on the tolerated failure probability. For a cost ratio $\lambda = {c_{FA}}/{c_{FR}}$, the corresponding failure-probability tolerance is $\alpha_\lambda = {1}/({1+\lambda}).$ Dimension $j$ is accepted whenever $\widehat{p}_{fail,j} \le \alpha_\lambda.$ Equivalently,
\begin{equation}
	\widehat{C}_{pk, j}
	\ge
	C_0 + k_\lambda SE(\widehat{C}_{pk, j}),
	\qquad
	k_\lambda = z_{1-\alpha_\lambda}.
	\label{eq:empirical_margin_accept}
\end{equation}

To examine how operational priorities influence approval outcomes, we evaluate decisions over a range of cost ratios $\lambda \in \{1,2,5,10,20,50\}$. For each value of $\lambda$, the analysis reports the numbers of accepted dimensions, rejected dimensions, and reclassifications relative to deterministic thresholding.

\begin{table*}[t]
	\caption{Dimension reclassification under risk-calibrated approval rules for the normal subset ($J_N=582$) and the non-normal subset ($J_{NN}=298$). Failure probabilities for the non-normal subset were estimated using bootstrap with $B=1000$ resamples. Here, A$\rightarrow$R denotes reclassification from accept to reject, and R$\rightarrow$A denotes reclassification from reject to accept.}
	\label{tab:reclassification_combined}
	\centering
	\begin{tabular*}{\textwidth}{@{\extracolsep{\fill}}ccccccccc}
		\hline
		& \multicolumn{4}{c}{Normal subset ($J_N=582$)} & \multicolumn{4}{c}{Non-normal subset ($J_{NN}=298$)} \\
		\cline{2-5} \cline{6-9}
		$\lambda$
		& Accepted & Rejected & A$\rightarrow$R & R$\rightarrow$A
		& Accepted & Rejected & A$\rightarrow$R & R$\rightarrow$A \\
		\hline
		Baseline & 291 & 291 & -- & -- & 105 & 193 & -- & -- \\
		$\lambda=1$  & 291 & 291 & 0  & 0 & 105 & 193 & 0  & 0 \\
		$\lambda=2$  & 285 & 297 & 6  & 0 & 101 & 197 & 4  & 0 \\
		$\lambda=5$  & 279 & 303 & 12 & 0 & 95  & 203 & 10 & 0 \\
		$\lambda=10$ & 267 & 315 & 24 & 0 & 88  & 210 & 17 & 0 \\
		$\lambda=20$ & 251 & 331 & 40 & 0 & 80  & 218 & 25 & 0 \\
		$\lambda=50$ & 245 & 337 & 46 & 0 & 75  & 223 & 30 & 0 \\
		\hline
	\end{tabular*}
\end{table*}

Table~\ref{tab:reclassification_combined} summarizes how approval decisions change when the deterministic threshold rule is replaced by the proposed risk-calibrated rule. As the cost ratio $\lambda$ increases, the approval margin widens and the number of accepted dimensions decreases monotonically in both subsets. Importantly, no reject-to-accept transitions occur, reflecting the conservative nature of the risk-calibrated rule.

A central empirical finding is that rule-dependent approval outcomes occur almost exclusively for dimensions whose estimated capability lies close to the approval threshold $C_0=1.33$. Dimensions with clearly insufficient capability remain rejected under all rules, whereas dimensions with very large capability estimates remain accepted regardless of the calibration method. Hence the practical effect of risk calibration is not to alter decisions for clearly capable or clearly incapable processes, but to stabilize decisions in the near-threshold regime where finite-sample uncertainty is most consequential.

The reclassification pattern is also substantively informative. The dimensions most affected by risk calibration are those that would have been marginally accepted under deterministic thresholding but are rejected once finite-sample uncertainty is taken into account. This is precisely the regime in which boundary instability is expected to matter most.

\subsection{Aggregate empirical risk under asymmetric cost}

Because the true capability status of each dimension is unknown, we construct an empirical proxy for the expected operational loss based on the estimated failure probability. For dimension $j$ and rule $\delta$, define the empirical risk score
\begin{equation}
	\widehat{EL}_j(\delta;\lambda)
	=
	\lambda\,\widehat{p}_{fail,j}\,\delta_j
	+
	(1-\widehat{p}_{fail,j})(1-\delta_j),
	\label{eq:empirical_el}
\end{equation}
where $\delta_j\in\{0,1\}$ denotes the observed decision. This proxy treats $\widehat{p}_{fail,j}$ as the estimated probability that dimension $j$ is truly below threshold, so that acceptance incurs expected cost proportional to $\lambda \widehat{p}_{fail,j}$ and rejection incurs expected cost proportional to $1-\widehat{p}_{fail,j}$.

Summing over all dimensions yields the aggregate empirical risk
\begin{equation}
	\widehat{EL}_{total}(\delta;\lambda)
	=
	\sum_{j=1}^{J}
	\widehat{EL}_j(\delta;\lambda).
	\label{eq:empirical_el_total}
\end{equation}

\begin{table*}[t]
	\caption{
		Aggregate empirical risk under deterministic threshold and cost-sensitive approval rules.
		For each cost ratio $\lambda=c_{FA}/c_{FR}$, the table reports the total empirical risk
		$\widehat{EL}_{total}$ for the deterministic threshold rule and for the corresponding risk-calibrated rule,
		together with the absolute risk difference
		$\Delta \widehat{EL}
		=
		\widehat{EL}_{total}(\text{deterministic};\lambda)
		-
		\widehat{EL}_{total}(\text{cost-sensitive};\lambda)$.
		Positive values of $\Delta$ therefore indicate risk reduction under the cost-sensitive rule.
		The percentage change is defined as
		$100 \times \Delta \widehat{EL} / \widehat{EL}_{total}(\text{deterministic};\lambda)$.
		For the normal subset ($J_{N}=582$), failure probabilities were estimated using the analytic normal approximation.
		For the non-normal subset ($J_{NN}=298$), failure probabilities were estimated by bootstrap resampling.
	}
	\label{tab:empirical_risk}
	\centering
	\begin{tabular*}{\textwidth}{@{\extracolsep{\fill}}lcccccccc}
		\toprule
		& \multicolumn{4}{c}{Normal subset ($J_N=582$)} & \multicolumn{4}{c}{Non-normal subset ($J_{NN}=298$)} \\
		\cmidrule(lr){2-5} \cmidrule(lr){6-9}
		$\lambda$
		& Deterministic
		& Cost-sensitive
		& $\Delta \widehat{EL}$
		& $\Delta (\%)$
		& Deterministic
		& Cost-sensitive
		& $\Delta \widehat{EL}$
		& $\Delta (\%)$\\
		\midrule
		1  & 16.155 & 16.155 & 0.000   & 0.0  & 12.997 & 12.931 & 0.066   & 0.5 \\
		2  & 23.001 & 21.835 & 1.166   & 5.1  & 17.910 & 16.481 & 1.429   & 8.0 \\
		5  & 43.540 & 32.456 & 11.085  & 25.5 & 32.649 & 19.657 & 12.992  & 39.8 \\
		10 & 77.772 & 41.989 & 35.784  & 46.0 & 57.214 & 21.668 & 35.546  & 62.1 \\
		20 & 146.237 & 50.874 & 95.363 & 65.2 & 106.344 & 23.258 & 83.086 & 78.1 \\
		50 & 351.630 & 58.484 & 293.145 & 83.4 & 253.734 & 25.956 & 227.778 & 89.8 \\
		\bottomrule
	\end{tabular*}
\end{table*}

Table~\ref{tab:empirical_risk} compares empirical risk under deterministic threshold and risk-calibrated approval rules. For $\lambda=1$, the two rules nearly coincide because symmetric costs imply essentially no additional approval margin. As $\lambda$ increases, the empirical risk under deterministic threshold approval rises rapidly, whereas the risk-calibrated rule introduces a finite-sample guard band that substantially reduces operational risk. 

Importantly, the empirical risk reduction is not driven by dimensions with extremely large capability estimates. Those dimensions are uniformly accepted under both deterministic threshold and risk-calibrated rules. Instead, the observed differences arise from dimensions near the approval threshold, for which finite-sample uncertainty can materially alter the approval decision.

The magnitude of this reduction is practically important. In the normal subset, the percentage risk reduction increases from $5.1\%$ at $\lambda=2$ to $83.4\%$ at $\lambda=50$. In the non-normal subset, the corresponding reduction increases from $8.0\%$ to $89.8\%$. These results suggest that the benefit of risk calibration becomes especially pronounced when false acceptance is substantially more costly than false rejection.

\begin{figure*}[t]
	\centering
	\subfloat[Distribution of estimated capability $\widehat{C}_{pk}$ across product dimensions.]{
		\label{fig:empirical_capability_distribution}
		\includegraphics[width=0.475\textwidth]{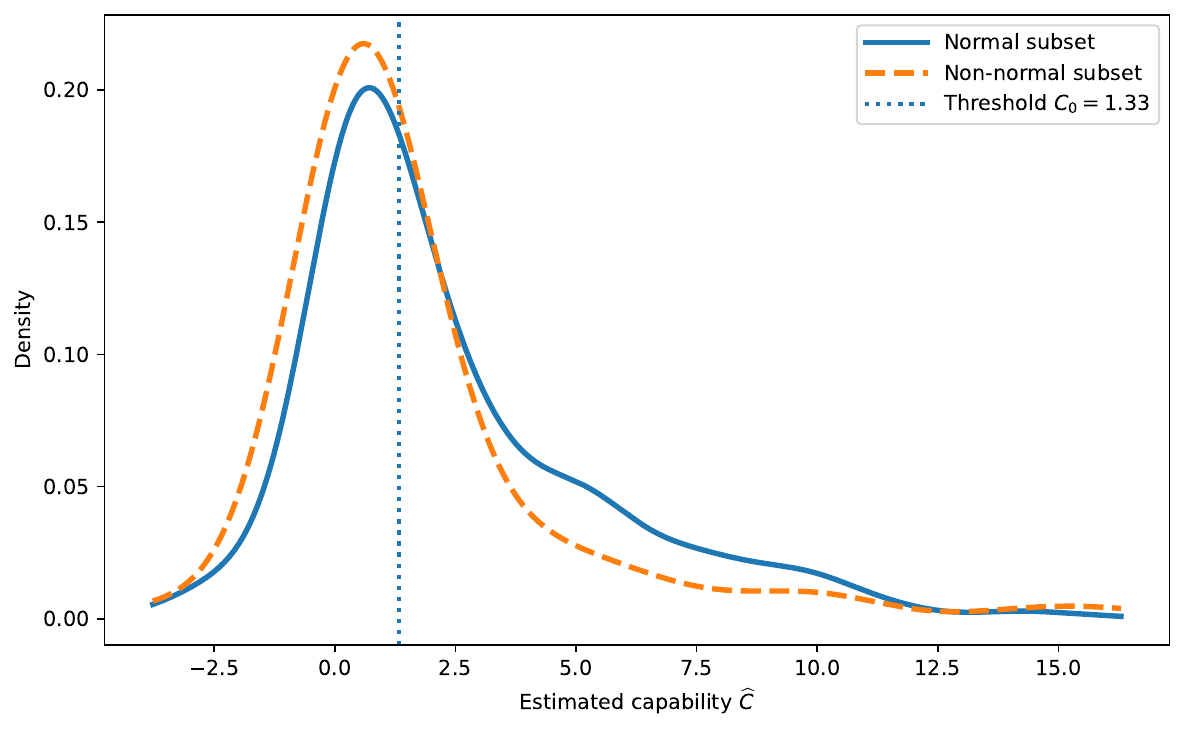}
	}
	\hspace{2pt}
	\subfloat[Empirical expected loss versus cost ratio $\lambda$ under deterministic threshold and risk-calibrated approval rules.]{
		\label{fig:empirical_expected_loss}
		\includegraphics[width=0.475\textwidth]{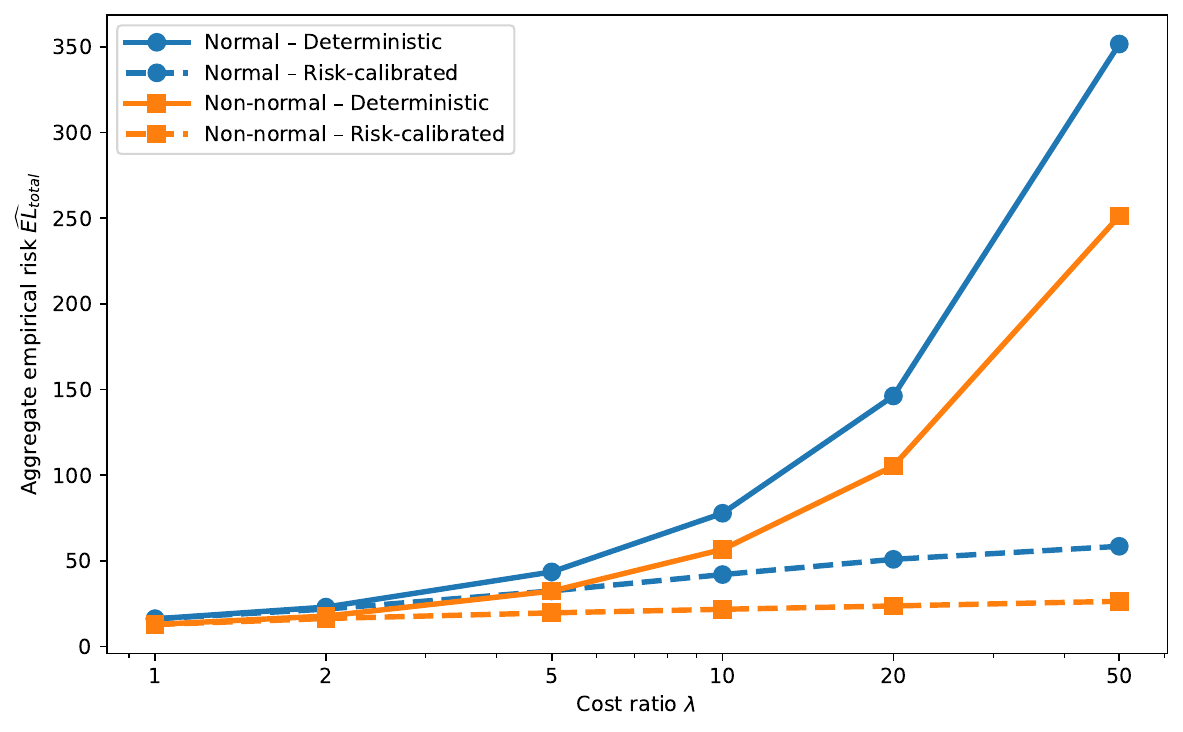}
	}
	\caption{
		Empirical characteristics of capability decisions in the industrial dataset.
		Panel (a) shows the distribution of estimated capability indices across all dimensions, separately for approximately normal and non-normal subsets. The vertical line indicates the approval threshold $C_0=1.33$.
		Panel (b) compares the aggregate empirical expected loss of deterministic threshold and risk-calibrated approval rules across cost ratios $\lambda$.
		The results illustrate how risk-calibrated decision rules can substantially reduce expected operational loss when false acceptance carries higher cost.
	}
	\label{fig:empirical_results}
\end{figure*}

Figure~\ref{fig:empirical_results} illustrates two complementary features of the empirical dataset. Panel~(a) shows substantial dispersion in capability estimates, with many dimensions concentrated near the approval threshold $C_0=1.33$. This concentration helps explain why approval decisions are sensitive to finite-sample uncertainty and why reclassification under uncertainty-adjusted rules is non-negligible.

Panel~(b) shows that deterministic thresholding leads to rapidly increasing empirical risk as the cost of false acceptance grows, whereas the risk-calibrated rule maintains substantially lower risk by incorporating an uncertainty-adjusted approval margin. The figure therefore complements Table~\ref{tab:empirical_risk} by visualizing the widening gap in decision performance as the asymmetry of operational loss becomes more severe.

Overall, the empirical results reinforce the theoretical analysis developed earlier in the paper. Explicitly accounting for finite-sample uncertainty leads to more robust, more conservative near-boundary decisions and more economically interpretable capability approval outcomes in practical manufacturing environments.

\section{Discussion and Conclusion}
\label{sec:discussion}

This paper develops a risk-calibrated framework for process capability approval under finite samples. Motivated by the boundary instability of deterministic thresholding, capability approval is formulated as a binary decision problem under asymmetric operational loss. This perspective shows that the conventional rule $\widehat{C}_{pk}\ge C_0$ implicitly corresponds to a particular risk calibration rather than an uncertainty-free classification.

Under a normal approximation to the sampling distribution of $\widehat{C}_{pk}$, the approval rule admits the unified guard-band representation \eqref{eq:k_family}, where the calibration constant $k$ determines the degree of protection against false acceptance. This representation unifies several commonly used approval procedures: deterministic thresholding corresponds to $k=0$, while lower confidence bound rules, posterior probability rules, and cost-sensitive approval rules arise from different choices of $k$. The analysis further shows that the deterministic rule $\widehat{C}_{pk}\ge C_0$ implies an asymptotic acceptance probability of $0.5$ when the true capability equals the approval threshold, indicating that it provides no explicit protection against erroneous approval in near-threshold regimes.

Introducing an uncertainty-adjusted margin yields a decision rule that becomes naturally more conservative when estimation uncertainty is large or when false acceptance carries greater operational cost. The calibration constant $k$ directly links operational priorities with statistical decision design: it can be specified through a tolerable failure probability $\alpha$ ($k=z_{1-\alpha}$) or through a false-accept/false-reject cost ratio $\lambda$ ($k=z_{1-\frac{1}{1+\lambda}}$).

Simulation experiments and the industrial case study demonstrate that risk calibration can substantially reduce expected operational loss when false acceptance is more costly than false rejection. The empirical analysis further shows that its impact is highly concentrated: it mainly affects dimensions whose estimated capability lies near the approval threshold, while leaving decisions unchanged for clearly capable or clearly incapable processes.

Although the closed-form margin rule is derived under an asymptotic normal approximation, the underlying decision principle is distribution-agnostic. When analytic standard errors are unreliable or the capability functional is non-normal, the same framework can be implemented using bootstrap estimation of the failure probability $P(C_{pk}^{true}<C_0\mid D)$. Taken together, these results suggest that uncertainty-aware capability approval provides a principled and practically useful refinement of conventional threshold-based decision making in manufacturing quality control.

\appendix
\numberwithin{equation}{section}
\numberwithin{figure}{section}
\section*{Appendix}
\section{Closed-form variance of $\widehat{C}_{pk}$ under normality}
\label{app:sigma}
The following derivation is based on a locally one-sided (active-constraint) regime, where the capability index is determined by either the upper or lower specification limit, but not both simultaneously. This corresponds to a centered or sufficiently off-centered process where one constraint dominates locally.

Assume the upper specification side is active, so that locally
\[
C_{pk}^{true}=g(\mu,\sigma)=\frac{\USL-\mu}{3\sigma}.
\]
Then
\[
\frac{\partial g}{\partial \mu}=-\frac{1}{3\sigma},
\qquad
\frac{\partial g}{\partial \sigma}
=
-\frac{\USL-\mu}{3\sigma^2}
=
-\frac{C_{pk}^{true}}{\sigma}.
\]
Hence
\[
\nabla g(\mu,\sigma)
=
\begin{pmatrix}
	-\frac{1}{3\sigma}\\[6pt]
	-\frac{C_{pk}^{true}}{\sigma}
\end{pmatrix}.
\]

Under normal sampling,
\[
\sqrt{n}
\begin{pmatrix}
	\overline{X}-\mu\\
	S-\sigma
\end{pmatrix}
\overset{d}{\to}
N\!\left(
0,
\begin{pmatrix}
	\sigma^2 & 0\\
	0 & \sigma^2/2
\end{pmatrix}
\right).
\]
Therefore,
\begin{align}
	\sigma_C^2
	&=
	\nabla g^\top \Sigma \nabla g
	\nonumber\\
	&=
	\left(-\frac{1}{3\sigma}\right)^2\sigma^2
	+
	\left(-\frac{C_{pk}^{true}}{\sigma}\right)^2\frac{\sigma^2}{2}
	\nonumber\\
	&=
	\frac19 + \frac{(C_{pk}^{true})^2}{2}.
	\label{eq:appendix_sigma_c}
\end{align}
Thus
\begin{equation}
	\sigma_C
	=
	\sqrt{\frac19+\frac{(C_{pk}^{true})^2}{2}}.
\end{equation}

\section{Bootstrap implementation of the probability rule}
\label{app:bootstrap}
When analytic standard error approximations are unreliable or when non-normal capability functionals are used, the posterior failure probability may be approximated by bootstrap.

Given the observed dataset $D$:
\begin{enumerate}
	\item Draw bootstrap resamples $D^{*(1)},\dots,D^{*(B_{boot})}$.
	\item Compute capability estimates $\widehat{C}_{pk}^{*(1)},\dots,\widehat{C}_{pk}^{*(B_{boot})}$.
	\item Estimate
			  \begin{equation}
			  	\widehat{p}_{fail}
			  	=
			  	\frac{1}{B_{boot}}\sum_{b=1}^{B_{boot}} \I(\widehat{C}_{pk}^{*(b)}<C_0).
			  \end{equation}
	\item Accept if
	\[
	\widehat{p}_{fail}\le \alpha.
	\]
\end{enumerate}

This procedure is distribution-agnostic and extends naturally to percentile-based capability indices for non-normal data.

\section*{DECLARATION}
\begin{description}
	\item[Funding:] This study received no external funding.
	
	\item[Conflicts of interest:] The authors declare no conflicts of interest.
	
	\item[Availability of data and material:] The empirical dataset is derived from anonymized manufacturing data. Processed data are available from the corresponding author upon reasonable request.
	
	\item[Code availability:] Simulation and analysis code are available from the corresponding author upon reasonable request.
	
	\item[Ethics approval:] Not applicable.
	
	\item[Consent for publication:] All authors approve the final manuscript.
\end{description}
\bibliographystyle{unsrtnat}
\bibliography{references}
\end{document}